# The effect of core formation on surface composition and planetary habitability

Brendan Dyck[12], Jon Wade[3], Richard Palin[3]

## Abstract

The melt productivity of a differentiated planet's mantle is primarily controlled by its iron content, which is itself approximated by the planet's core mass fraction (CMF). Here we show that estimates of an exo-planet's CMF allows robust predictions of the thickness, composition and mineralogy of the derivative crust. These predicted crustal compositions allow constraints to be placed on volatile cycling between surface and the deep planetary interior, with implications for the evolution of habitable planetary surfaces. Planets with large, terrestrial-like, CMFs ($\geq 0.32$) will exhibit thin crusts that are inefficient at transporting surface water and other volatiles into the underlying mantle. By contrast, rocky planets with smaller CMFs ($\leq 0.24$) and higher, Mars-like, mantle iron contents will develop thick crusts capable of stabilizing hydrous minerals, which can effectively sequester volatiles into planetary interiors and act to remove surface water over timescales relevant to evolution. The extent of core formation has profound consequences for the subsequent planetary surface environment and may provide additional constraints in the hunt for habitable, Earth-like exo-planets.

[1] Corresponding author Brendan.dyck@ubc.ca
[2] Department of Earth, Environmental and Geographic Sciences, University of British Columbia, Kelowna, V1V 1V7, Canada

[3] Department of Earth Sciences, University of Oxford, Oxford, OX1 3AN, United Kingdom

## INTRODUCTION

The planets Mercury, Earth and Mars are all broadly chondritic in their ratios of major refractory elements, but possess silicate-mantle iron contents that increase with heliocentric distance (Trønnes et al. 2019). This compositional variation is a result of the differing relative proportions of iron that entered their cores during early differentiation. The extent of planetary core formation as a function of total planet mass – the core mass fraction (CMF) - thus reflects the oxidation gradient present in the proto-planetary disc and the increasing contribution of oxidized, outer solar system material to planetary feedstocks (O'Brien et al. 2018). The iron content of a rocky planet's silicate mantle is determined by the prevailing oxygen fugacity of core formation (Frost & McCammon 2008), such that more oxidizing conditions lead to higher mantle iron contents (Wade & Wood 2005). Oxidation gradients have been observed around other main sequence stars (Putirka & Rarick 2019), and similar gradients in mantle iron contents are thus expected in other planetary systems possessing rocky differentiated planets (Agol et al. 2020) (Figure 1). Consequently, even if each rocky body in a multi-planetary system forms from similar precursor material, variations in their core mass fraction will generate silicate mantles and derivative surface crusts that exhibit distinct compositional and petrophysical differences. Hence, variations in CMF may have a disproportionate role in determining a planet's geological evolution and its future habitability.

Excluding any subsequent loss of mantle, mass-balance constraints for a solar chondritic abundance of elements predict a maximum CMF for the terrestrial planets of ~0.34. This value decreases with heliocentric distance away from the Sun, reflecting the greater contribution of oxidized components originating from beyond the snow-line (Figure 1) to the growing planet (Lichtenberg et al. 2020). Consequently, planetary mantles become increasingly iron-rich with proximity to the snow-line. To explore how the extent of core formation influences the thickness and chemical make-up of a planet's crust, we performed petrological forward modelling using a Gibbs free energy minimization procedure that simulates adiabatic mantle decompression melting and crust production in chondritic planets with CMFs between 0.34 and 0.16. In contrast with previous experimental and modelling studies that evaluate the influence of bulk planet composition on mantle mineralogy and crust production (e.g. Putirka & Rarick 2019), we hold bulk planet composition constant varying only the distribution of iron between core and mantle. We show that planets with large CMFs (≥0.32) and iron-poor mantles (about 0–4 wt. % FeO; cf. Mercury) generate thin, feldspar-rich crusts that can carry relatively little water during burial or subduction into the planet's interior.

A decreasing CMF and increasing mantle iron content (up to 25 wt. % FeO; cf. Mars) results in a concomitant increase in the thickness of the crust generated and contains abundant olivine and pyroxene. Unlike the mineralogy of reduced planetary crusts, these minerals readily weather at planetary surfaces and in doing so incorporate volatiles and stabilize hydrous minerals during metamorphism that can transport water from the hydrosphere to mantle depths (Wade et al. 2017). As a planet ages and cools, the influence of CMF on both crust thickness and hydration capacity become more pronounced, underscoring the sustained role of core formation in setting the surface evolution, and subsequent habitability, of rocky planets.

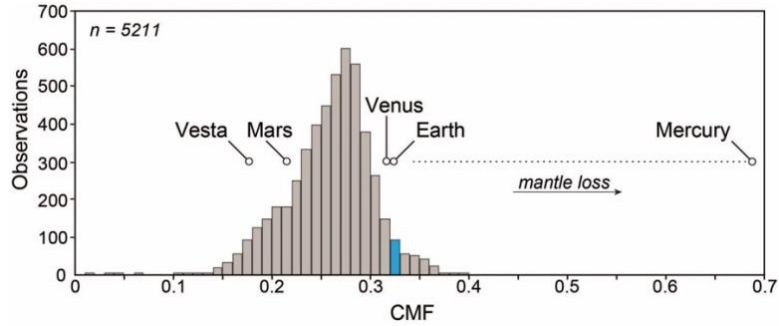

**Figure 1.** Histogram of core-mass fraction (CMF) estimates for extrasolar rocky planets orbiting FGK stars. Data for 5211 stars from the Hypatia Catalog (Hinkel et al. 2014), with Lodders et al. (2009) as the solar reference (after Unterborn & Panero 2019). Bin width is 0.01. A solar composition predicts a rocky planet with a CMF ~0.33 (blue bin). The diversity of CMF within our solar system reflects an oxidation gradient during planetary accretion.

## METHOD

To model the formation of juvenile basaltic crust, we adopt the approach of Weller et al. (2019) and assume that mantle decompression melting occurs isentropically, that thermal equilibrium is maintained between solid and melt phases, and that melt is efficiently extracted at low melt fractions (see detailed method description in Supporting Material). We considered mantle potential temperatures ($T_P$) of 1,500 °C and 1,700 °C. With near-surface peridotite liquidus temperatures constrained to ~1,750–1,800 °C (Takahashi et al. 1993), convective cooling of the upper mantle sets a realistic upper bound of ~1,750 °C to the operation of near-surface adiabatic decompression melting (Elkins-Tanton et al. 2005). However, for rocky planets with similar masses to those in our solar system (0.055–1 $M_\oplus$, where $M_\oplus$ = mass of Earth), the prevailing upper mantle $T_P$ at the time of substantial juvenile lithosphere formation would have been around 1,500 ± 200 °C (Stevenson, 2003). Earth has experienced some form of basaltic (oceanic) crust generation since at least the Mesoarchean (ca. 2.8–3.2 Ga), when mantle $T_P$ reached a maximum of ~1,675 °C (Korenaga 2008). Continued juvenile crust formation has occurred during planetary cooling to the present day, where the ambient mantle $T_P$ is ~1,350 °C. By contrast, the majority of Mars' basaltic crust formed at a mantle $T_P$ of ~1,500 °C (Lessel & Putirka 2015). The rate of secular cooling of terrestrial planetary mantles from a supra-liquidus (>1,800 °C) to subsolidus (<1,400 °C) state depends on planetary mass, surface heat loss, and degree of mantle melting (Korenaga 2008); thus, we utilize a mantle $T_P$ of 1,500 °C as a common reference temperature to compare which factors affect the efficiency of terrestrial crust generation.

## RESULTS

### Crust Thickness Estimates

To contextualize our results, we calculated crust thicknesses as a function of mantle iron content for Mars-Earth-, and Super Earth-sized planets where $g$ is ~3.7, ~9.8, and ~14 m/s$^2$, respectively (Figure 2). An increase in mantle iron content lowers the melting temperature of silicate mantle, resulting in a larger melt volume generated over a broader range of pressures (Figure 3). At a $T_P$ of 1,700 °C, increasing mantle FeO from 0.5 to 25 wt. % (CMFs of 0.34 and 0.16, respectively) more than doubles the crustal thickness (Table S3). Lowering the $T_P$ to 1,500 °C results in an eight-fold increase in the thickness of planetary crust over the same range of mantle FeO. For a Mars-sized planet with a large iron core (CMF = 0.34) and low-iron mantle (0.5 wt. % FeO), our model predicts a crust thickness of 3.7 ± 1.4 km (Table S3). If the same planet accreted under more oxidizing conditions and evolved a mantle with elevated FeO (25 wt. %), corresponding to a CMF of 0.16, it would instead form a crust that is 65.4 ± 9.8 km thick (Table S3).

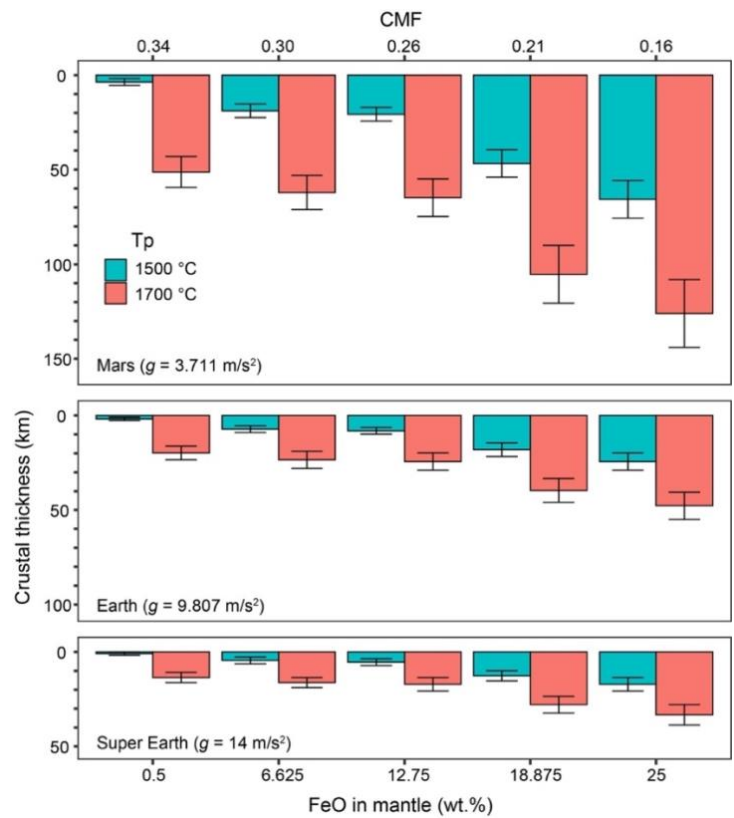

**Figure 2.** Crust thickness a function of mantle potential temperature ($T_P$) and mantle FeO (with corresponding CMF) for Mars-, Earth-, and Super Earth-like planets. Data are shown for $T_P$ = 1500 °C and 1700 °C, and error bars are given at 2σ.

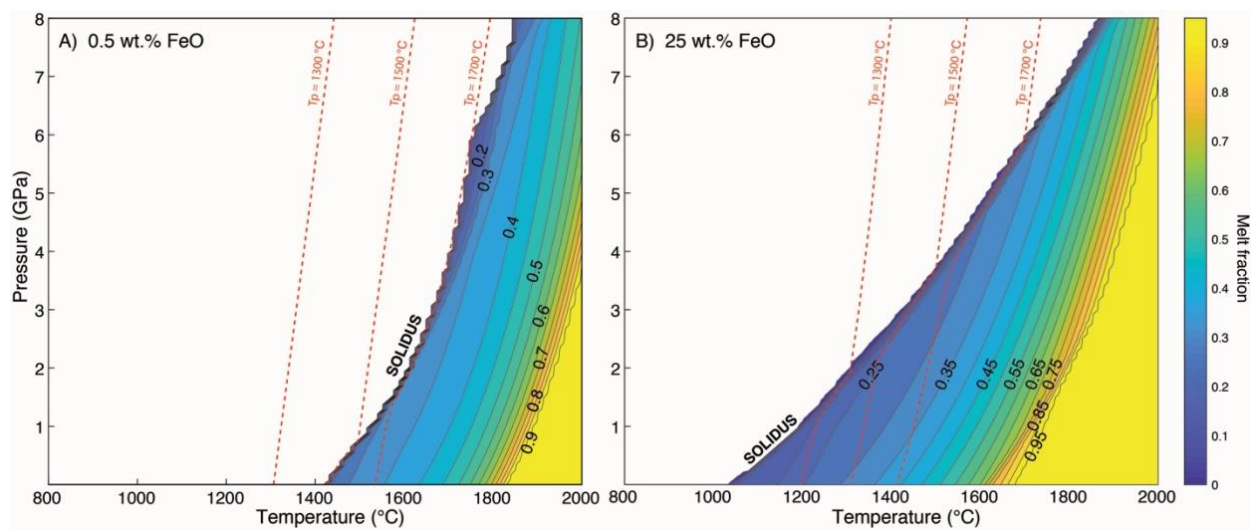

**Figure 3.** Pressure–temperature phase diagrams showing isentropic melting relationships for silicate mantles with A: low (0.5 wt. %) and B: high (25 wt. %) FeO contents. Dashed red lines represent isentropes for adiabatic decompression melting pathways for mantle potential temperatures ($T_P$) of 1300 °C, 1500 °C, and 1700 °C.

*Composition of Juvenile Crust*

Further to its influence on the volume of crust generated, the extent of planetary core formation also determines the chemistry and mineralogy of planetary lithospheres. Due to suppression of the peridotite solidus with increasing maficity (Takahashi et al. 1993; Tuft et al. 2005), planets with higher mantle iron contents undergo incipient mantle melting at higher pressures and generate crust across a wider temperature range than their low-mantle-iron counterparts (see Figure 3). As recently noted in Lambart et al. (2016) and Putirka and Rarick (2019), planetary crust production is also sensitive to the modal proportion of clinopyroxene in the mantle. However, since the degree of core formation has little influence on modal mantle mineralogy owing to the solid-solution of Fe-Mg in the four major mantle phases (olivine, orthopyroxene, clinopyroxene, melt), the effect of varying bulk planet composition is largely independent from the findings reported here.

At any given $T_P$, increasing the amount of iron in planetary mantles produces a more primitive, olivine-rich, basaltic crust with lower Al/Si and elevated iron, akin to Martian crust (Figure S1) (McSween et al. 2009). However, increasing mantle iron content also reduces both the mantle's solidus temperature and the viscosity of the resultant basaltic magmas (Tuft et al. 2005; Sehlke & Whittington 2016), driving the continued generation of juvenile crust as a planet ages and cools. At the high-iron end of our simulations (CMF of 0.16) mantle decompression melting continues at shallow depths even as $T_P$ cools below 1100 °C (Figure 3). Hence volcanism on planets with higher mantle iron contents generates surfaces that are dominated by long-lived flood basalt fields and shield volcanoes (Hauber et al. 2011). Erupted or emplaced onto a wet planetary surface, these evolved, iron bearing magmas may be assumed to efficiently hydrate, forming hydrous minerals and consuming hydrospheric water in the process (Wade et al. 2017).

## DISCUSSION

### Fate of Surface Volatiles

Water cycles across planetary surfaces via the volumetrically abundant hydrous minerals of the serpentine and amphibole groups, the stability of which varies with the composition of juvenile crust. Serpentinization – the reaction of olivine with water at or near a planet's surface to produce magnetite and phyllosilicate minerals of the serpentine group – is a primary driver of hydrosphere deprotonation. This weathering process is widespread on the Earth's surface and is known to occur on Mars, with phyllosilicates identified *in-situ* (Amador et al. 2018) and from orbit (Carter et al. 2013). The dissociation of water during silicate-mineral hydration reactions may also be responsible for the formation of the carriers of the Martian remnant magnetic field (Lillis et al. 2008), magnetite and maghemite. Serpentine minerals are, however, unstable above ~600 °C and, in the absence of subduction, will not transport volatiles to mantle depths on either Earth or Mars. Rather, it is the stability of amphibole—a common constituent of metamorphosed basaltic crust—that drives sequestration of volatiles in the upper mantle (Smye et al. 2017). In addition to the stoichiometric hydroxyl and fluorine, similarity between the ionic radii of alkali metals Na and K, permits the transport of halogens and other volatiles into the deep interior. Amphibole has therefore likely formed an integral part of the deep-Earth water cycle since the onset of plate tectonics several billion years ago, replenishing the upper mantle with volatiles (Smye et al. 2017). With an upper stability limit >1000 °C, amphibole may also play a critical role in the volatile cycle and the progressive oxidation of the upper mantles of planets that exhibit a stagnant-lid tectonic regime (Wade et al. 2017). Under this scenario, hydrated basaltic crust may be buried by repeated lava flows, transformed to amphibolite, and gravitationally drip or delaminate into the underlying convective mantle. Such repeated crustal over-plating is facilitated by the high iron content of the primitive mantle which both lowers the mantle solidus and decreases the melt viscosity. This is exacerbated by the transport of volatiles to depth which act to decrease further both the mantle solidus and the viscosity of the resultant lavas.

Calculated mineral modes for crusts generated on low-, medium-, and high-mantle-iron Mars-sized planets as a function of depth within the interior are show in Figure 4. With increasing mantle iron content, the erupted basalt compositions are both more olivine-rich – leading to an increased serpentinization potential (Oze & Sharma 2005) – and, as a consequence of the increased crustal thickness, more readily enable amphibole to stabilize in the lower crust. Importantly, a crust generated from a low-iron mantle is olivine-normative, but too thin to form hornblendic amphibole, which stabilizes at depths of ~18–20 km. Furthermore, higher conductive geotherms within the markedly thin crust generated on planets with large CMFs (>0.32) would promote the near-surface decomposition of hydrous minerals, effectively restricting the cycling of volatiles to the outermost planetary layer where it can be stripped away by solar radiation (Jakosky et al. 2017). Our results predict that a Mercury-like planet, with a low mantle iron content and high CMF, would develop a thin, anhydrous crust incapable of maintaining a water cycle that extended beyond the shallowest levels of the lithosphere. Spectroscopic analyses of Mercury's surface lavas provide support for our model results (Namur & Charlier 2017).

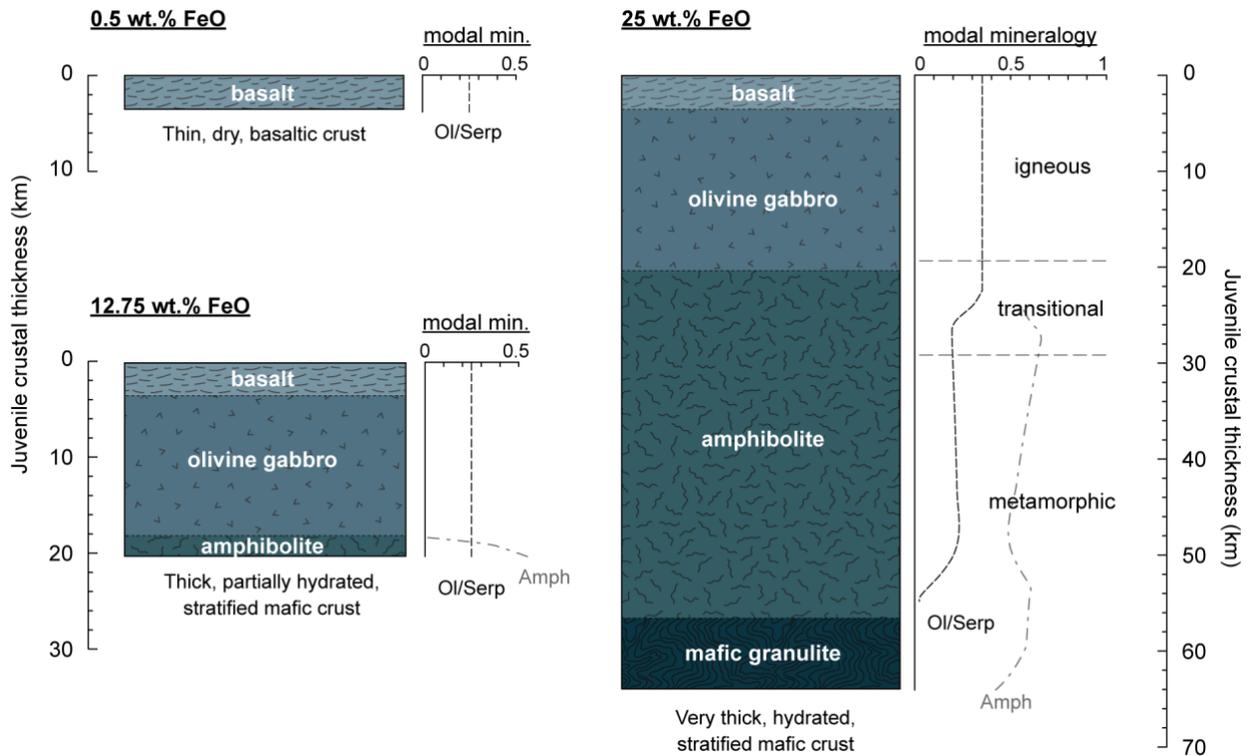

**Figure 4.** Schematic sections showing juvenile crustal thickness, lithology, and modal mineralogy for a Mars-sized planet with different mantle FeO contents.

## CONCLUSIONS

Our modeling shows that retention of water on a planet's surface for geological timescales is principally related to the silicate mantle's iron content, itself a result of the heliocentric distance and the stochastic processes attendant with planetary core formation. By determining the silicate mantle's iron content, CMF exerts a first-order control on the thickness of resultant crust formed by the adiabatic melting of a rocky planet's mantle (Figure 5). Planets with broadly chondritic bulk compositions that possess smaller CMFs will exhibit more extensive volcanism leading to relatively thicker crusts capable of stabilizing amphiboles and, consequently, have a higher potential for crustal hydration. Numerical modeling of lithospheric behavior on the Archean Earth indicates that thicker crusts are more buoyant and difficult to subduct (Van Hunen & Moyen 2012), promoting a stagnant-lid geodynamic regime on such planets. Burial of hydrated crustal materials by repeated volcanic resurfacing events and transport into the upper mantle by Rayleigh-Taylor type lithospheric drips results in the continuous sequestration of volatiles into the planetary interior, with few avenues for recycling. The CMF, and conditions of planetary scale differentiation and the resultant iron content of the mantle that it represents, may therefore be considered a key determinant a planet's subsequent evolution and habitability.

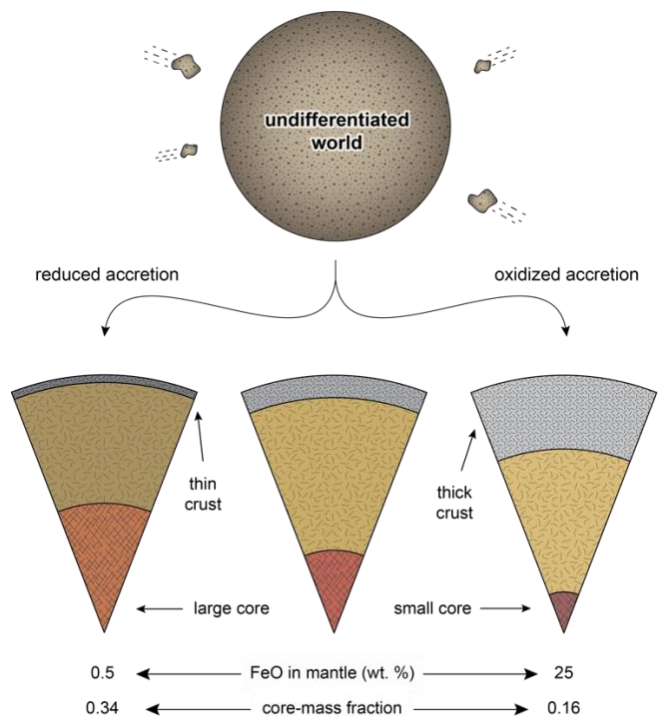

**Figure 5.** Schematic showing the scenarios we investigated. The redox conditions of core formation play a formative role in determining crustal thickness and composition.

### Acknowledgments

This work was supported by the Natural Sciences and Engineering Research Council of Canada (RGPIN-04248). Author Contributions: B.D. conceived the idea in discussion with J.W. and performed all calculations. Keith Putirka is thanked for a helpful review. All authors contributed to writing the manuscript. All data is available in the main text or the supplementary materials.

**Supporting Materials**

*Extended Methods*

Our petrological modeling was performed in three steps. First, we determined suitable 'bulk-silicate body' compositions for planetary mantles by using the Martian mantle composition (Yoshizaki & McDonough 2020) as a starting point and reducing its Fe content to simulate different degrees of core growth during initial differentiation. Five mantle compositions were generated in this way, which correspond to CMFs of 0.34, 0.30, 0.26, 0.21, and 0.16, respectively (Table S1). Second, we simulated the generation of juvenile crust for planets with varying CMF by partially melting the silicate mantle compositions of part (1), using a peridotite melt model in the Theriak-Domino program (de Capitani & Petrakakis 2010). Third, with the melt compositions generated in part (2) we calculated normative mineral modes for juvenile igneous crust over the predicted range from the surface (25 °C) to 500 °C and the modal abundance of metamorphic phases from 500–1200 °C.

*Petrological modelling procedure*

All of the phase equilibria calculations were made using Theriak-Domino (de Capitani & Petrakakis 2010). Mantle melting relations were calculated in the $Na_2O–CaO–FeO–MgO–Al_2O_3–SiO_2–O_2–Cr_2O_3$ (NCFMASOCr) compositional system using the activity–composition ($a$–$X$) relations of (Jennings & Holland 2015), and the internally consistent thermodynamic dataset ds-62 (Holland & Powell 2011). For internal consistency and ease of comparison, calculations for all rock types used an average terrestrial mantle molar bulk-rock $X(Fe^{3+}) = Fe_2O_3/(Fe_2O_3 + FeO)$ value of 0.1 (Davis et al. 2009). The stable mineral assemblage in the metamorphic crust was calculated in the $MnO–Na_2O–CaO–K_2O–FeO–MgO–Al_2O_3–SiO_2–H_2O–TiO_2–O_2$ (MnNCKFMASHTO) system using the internally consistent thermodynamic dataset ds-55 of (Holland & Powell 1998) (updated to August 2004) and $a$–$X$ relations for melt, garnet, biotite and ilmenite, feldspar, orthopyroxene and spinel, amphibole and clinopyroxene after (Dyck et al. 2020). $H_2O$ was set to saturate the system over the entire range of modelled conditions. To maintain petrological consistency between the generated crust compositions and the bulk compositions used when calculating crust mineralogy, we set MnO and $TiO_2$ to be 0.01 mols and $K_2O$ 0.02 mols.

To simulate the generation of juvenile crust by adiabatic decompression we modelled the melting behavior of peridotite mantle along pressure–temperature paths with constant entropy (isentropes) (Weller et al. 2019). We account for chemical fractionation during melting by extracting melt at ~10 °C intervals, from the temperature at which a given isentrope intersects the solidus ($Tp$) and to the temperature at which the isentrope intersects the surface (Figure S2). As the removal of melt reduces the absolute value of entropy in the system, we conserve mass and define our isentropic model system to include both the mantle and the extracted melt. Consequently, batch (incremental removal of melt) and equilibrium melting (where melt is not removed) follows the same isentrope.

To simulate the effect of a small degree of residual melt remaining in the melt source, we removed 90 vol. % of the melt phase present at each melt interval. With an average of 0.01–0.02 melt mode (1–2 vol. % of system) being removed at each interval, our procedure closely approximates the adiabatic melting behavior of the terrestrial mantle (Mckenzie & Bickle 1988).

The composition and thickness of the juvenile crust was calculated for each bulk composition (Tables S2 & S3) following the procedure outlined in (Weller et al. 2019). Crust thickness is the sum of the product between layer height ($h$) and melt fraction (F) for each calculation interval, where $h$ (km) = $P$ (bars) / 10 $\rho$ (g·cm$^{-3}$) $g$ (m·s$^2$), and $\rho$, which is a function of the mineral assemblage and compositions, was independently solved for in the phase equilibria. Using the composition of melt (juvenile crust) generated over each interval and the relative contribution that each interval made to cumulative crustal thickness, a weighted average crust composition was calculated that accounts for the greater volume of melt produced nearer the solidus as a given column of mantle rises from the solidus to the surface (Table S2).

We benchmarked our model for a range of mantle FeO contents by calculating the thickness and composition of the basaltic (oceanic) crusts for Earth and Mars using average upper mantle compositions KLB1 (Earth (Davis et al. 2009)) and Bulk Silicate Mars (Dreibus & Wanke 1985) as our starting mantle compositions. The resulting crustal thicknesses (Table S4) closely match the reference values of Earth's present-day oceanic crust (~7 km) and the thickness estimates for Martian crust (35–70 km (Tenzer et al. 2015)). Similarly, the modelled composition of Earth's crust closely resembles global average mid-ocean ridge basalts, whereas the modelled Martian crust approximate the shergottite meteorites.

Uncertainty on the absolute positions of assemblage field boundaries in pressure–temperature space generally do not exceed ± 0.5 kbar and ± 30 °C for the simple NCFMASOCr model system and ± 1 kbar and ± 50 °C for the extended MnNCKFMASHTO model system at the 2σ level (Powell & Holland 2008). As this variation is largely a function of propagated uncertainty on end-member thermodynamic properties within the internally consistent data set and, as all phase equilibria calculations were performed using the same dataset and activity–composition relations, calculated phase equilibria are expected to be relatively accurate to within ± 0.2 kbar and ± 10–15 °C(Powell & Holland 2008). We report crustal thickness with 2σ absolute uncertainty, which is calculated by propagating an uncertainty of ± 30 °C in $T_P$ through our model.

We adopt the 'Early Martian' d$T$/d$P$ gradient used in (Wade et al. 2017) and calculate stable mineral assemblages for juvenile igneous crust over the predicted range from the surface (25 °C) to 1200 °C. As the metamorphism is kinetically limited at lower temperatures and the activity-composition models are not robust below ~500° C, we report normative (igneous) mineral modes for temperatures <500 °C (Hollocher 2011) and the modal abundance of metamorphic phases for temperatures between 500–1200 °C.

Table S1: Bulk silicate mantle compositions used for petrological modeling (wt. % oxide). CMF = core mass fraction.

| CMF | FeO | SiO$_2$ | Al$_2$O$_3$ | Cr$_2$O$_3$ | MgO | CaO | Na$_2$O |
|------|-------|---------|-------------|-------------|-------|------|---------|
| 0.34 | 0.50  | 38.89   | 5.73        | 0.50        | 48.80 | 4.57 | 0.76    |
| 0.30 | 6.63  | 38.39   | 5.20        | 0.45        | 44.26 | 4.14 | 0.69    |
| 0.26 | 12.75 | 37.90   | 4.67        | 0.40        | 39.73 | 3.72 | 0.62    |
| 0.21 | 18.88 | 37.40   | 4.13        | 0.36        | 35.20 | 3.30 | 0.55    |
| 0.16 | 25.00 | 36.91   | 3.60        | 0.31        | 30.67 | 2.87 | 0.48    |

Table S2. Modeled average juvenile crust composition (wt. % oxide) and normative mineralogy as a function of mantle potential temperature ($T_P$) and core mass fraction (CMF). Mg# = molar Mg/(Mg + Fe$^{2+}$). Ol = olivine; Cpx = clinopyroxene; Pl = plagioclase; other = other minerals, including larnite and nepheline. CIPW norms were calculated following the procedure outlined by Hollocher (2011).

| | Mantle FeO (wt. %) | CMF | Composition | | | | | | | | CIPW norms | | | |
|---|---|---|---|---|---|---|---|---|---|---|---|---|---|---|
| | | | SiO$_2$ | Al$_2$O$_3$ | Na$_2$O | CaO | FeO | MgO | Cr$_2$O$_3$ | Mg# | Ol | Cpx | Pl | Other |
| $T_P$ = 1500 °C | 0.50 | 0.34 | 39.90 | 23.80 | 1.19 | 18.28 | 0.71 | 16.12 | 0.00 | 0.98 | 25 | 0 | 61 | 14 |
| | 6.63 | 0.30 | 45.81 | 18.48 | 4.47 | 12.11 | 6.23 | 12.81 | 0.09 | 0.79 | 20 | 21 | 38 | 21 |
| | 12.75 | 0.26 | 44.69 | 15.81 | 3.95 | 11.19 | 12.39 | 11.88 | 0.09 | 0.63 | 25 | 23 | 34 | 18 |
| | 18.88 | 0.21 | 44.18 | 13.79 | 3.61 | 10.97 | 16.14 | 11.24 | 0.07 | 0.55 | 27 | 26 | 30 | 17 |
| | 25.00 | 0.16 | 42.27 | 10.09 | 2.06 | 12.06 | 23.14 | 10.32 | 0.06 | 0.44 | 33 | 32 | 22 | 13 |
| $T_P$ = 1700 °C | 0.50 | 0.34 | 42.80 | 19.87 | 2.21 | 15.09 | 0.66 | 19.36 | 0.00 | 0.98 | 30 | 5 | 47 | 18 |
| | 6.63 | 0.30 | 42.94 | 15.66 | 2.42 | 12.47 | 8.44 | 17.93 | 0.13 | 0.79 | 33 | 15 | 35 | 17 |
| | 12.75 | 0.26 | 44.06 | 14.64 | 3.57 | 9.91 | 11.73 | 15.94 | 0.14 | 0.71 | 32 | 19 | 31 | 18 |
| | 18.88 | 0.21 | 42.39 | 13.10 | 2.35 | 11.51 | 15.44 | 15.08 | 0.13 | 0.64 | 35 | 21 | 29 | 15 |
| | 25.00 | 0.16 | 41.47 | 9.02 | 1.71 | 10.31 | 23.59 | 13.79 | 0.10 | 0.51 | 42 | 27 | 21 | 10 |

Table S3. Calculated thicknesses of juvenile crust on planets of various sizes as a function of mantle potential temperature ($T_P$) and composition. CMF = core mass fraction

| Planet and associated gravity ($g$) (m²/s) | Mantle wt. % FeO | CMF | $T_P$ = 1500 °C | | $T_P$ = 1700 °C | |
|---|---|---|---|---|---|---|
| | | | Crust thickness (km) | ±30 °C uncertainty (km) | Crust thickness (km) | ±30 °C uncertainty (km) |
| Mars ($g$ = 3.701) | 0.50 | 0.34 | 3.7 | 1.4 | 51.4 | 7.9 |
| | 6.63 | 0.30 | 18.7 | 3.6 | 61.8 | 9.3 |
| | 12.75 | 0.26 | 20.3 | 3.8 | 64.4 | 9.7 |
| | 18.88 | 0.21 | 47.0 | 7.4 | 104.9 | 15.1 |
| | 25.00 | 0.16 | 65.4 | 9.8 | 125.8 | 17.9 |
| Earth ($g$ = 9.807) | 0.50 | 0.34 | 1.4 | 1.1 | 19.4 | 3.7 |
| | 6.63 | 0.30 | 7.1 | 2.0 | 23.3 | 4.2 |
| | 12.75 | 0.26 | 7.7 | 2.1 | 24.3 | 4.3 |
| | 18.88 | 0.21 | 17.6 | 3.4 | 39.6 | 6.4 |
| | 25.00 | 0.16 | 24.7 | 4.4 | 47.5 | 7.4 |
| Super Earth ($g$ = 14) | 0.50 | 0.34 | 1.0 | 1.1 | 13.6 | 2.9 |
| | 6.63 | 0.30 | 4.9 | 1.7 | 16.3 | 3.3 |
| | 12.75 | 0.26 | 5.4 | 1.8 | 17.0 | 3.3 |
| | 18.88 | 0.21 | 12.4 | 2.7 | 27.7 | 4.8 |
| | 25.00 | 0.16 | 17.3 | 3.4 | 33.3 | 5.5 |

Table S4. Benchmark results for petrological modeling of crust generation for Earth and Mars. Initial average mantle compositions (wt. % oxide) and modeled compositions (wt. % oxide) and thicknesses (km) of crust on Earth and Mars. Mg# = molar Mg/(Mg + Fe$^{2+}$). $T_P$ = mantle potential temperature. Thicknesses are shown with a 2-sigma uncertainty. Composition of KLB1 terrestrial peridotite is from Davis et al. (2009). Composition of bulk-silicate Mars (BSM) is from Yoshizaki and McDonough (2020). Normal MORB composition and thickness after Gale et al, (2013) and White et al. (1992), respectively. Martian crustal thickness estimate after Tenzer et al. (2015). Martian composition (Fastball) after Squyres et al. (2007) with $T_P$*estimate after Filiberto et al. (2010).

| | Planet | Composition (wt. % oxide) | | | | | | | | $T_P$ (°C) | Thickness (km) |
|---|---|---|---|---|---|---|---|---|---|---|---|
| | | $SiO_2$ | $Al_2O_3$ | $Na_2O$ | CaO | FeO | MgO | $Cr_2O_3$ | Mg# | | |
| **Initial mantle** | Earth (KLB1) | 45.18 | 4.47 | 0.36 | 3.56 | 8.08 | 37.95 | 0.39 | 0.89 | | |
| | Mars (BSM) | 44.47 | 2.94 | 0.51 | 2.43 | 17.89 | 30.16 | 0.77 | 0.75 | | |
| **Modelled crust** | Earth | 48.27 | 17.31 | 2.96 | 12.51 | 7.88 | 10.91 | 0.17 | 0.71 | 1300 | 7.7 ± 2.1 |
| | Mars | 49.92 | 10.31 | 3.61 | 8.20 | 18.30 | 9.38 | 0.28 | 0.48 | 1300 | 23.5 ± 3.2 |
| | Mars | 48.26 | 8.57 | 2.90 | 8.47 | 20.64 | 10.82 | 0.35 | 0.48 | 1400 | 41.0 ± 3.8 |
| | Mars | 46.97 | 7.06 | 2.42 | 8.05 | 22.41 | 12.66 | 0.43 | 0.50 | 1500 | 66.6 ± 4.5 |
| **Observed crust** | Earth (NMORB) | 51.68 | 15.51 | 2.90 | 11.63 | 10.06 | 7.95 | 0.27 | 0.44 | 1300–1350 | 7.1 ± 0.8 |
| | Mars (F07) | 49.46 | 8.57 | 2.57 | 6.33 | 19.43 | 13.10 | 0.53 | 0.41 | 1480–1530* | 35–70 |

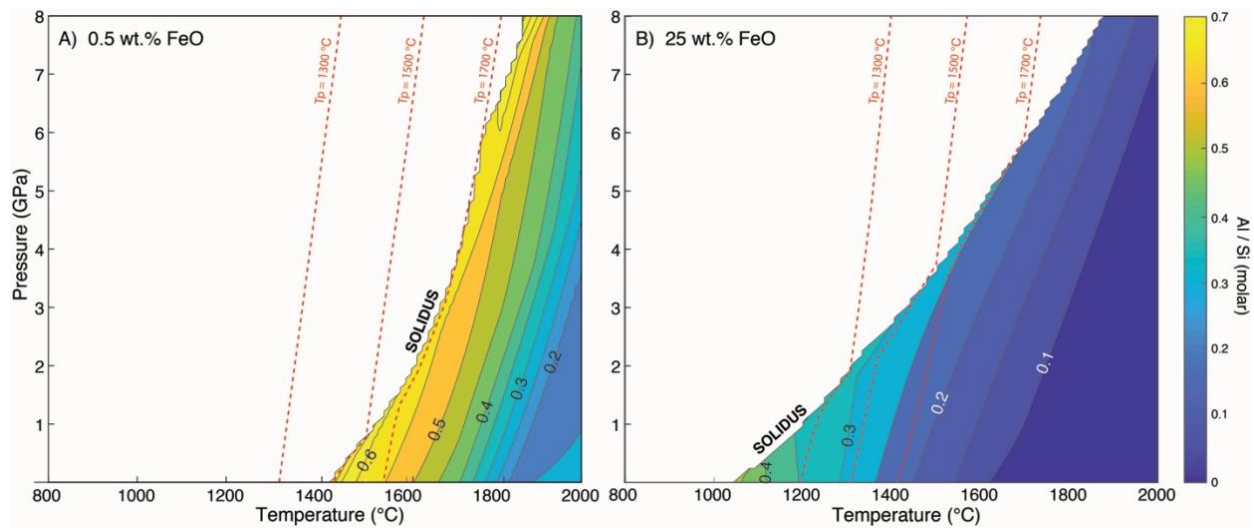

**Figure S1.** Pressure–temperature phase diagrams showing the atomic Al/Si composition of melt generated during adiabatic decompression for silicate mantles with A: low (0.5 wt. %) and B: high (25 wt. %) FeO contents. Dashed red lines represent isentropes for adiabatic decompression melting pathways for mantle potential temperatures ($T_P$) of 1300 °C, 1500 °C, and 1700 °C.

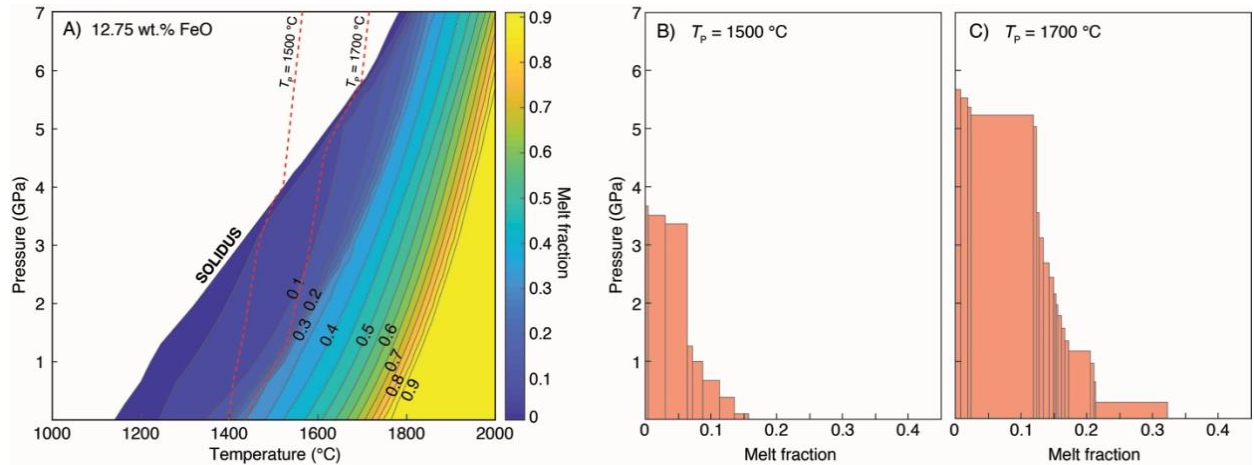

**Figure S2.** Petrological model for fractional mantle melting and crust formation on a Mars-sized body. (A) Pressure–temperature phase diagram showing the solidus and calculated suprasolidus melt fractions. Red dashed lines represent isentropes for mantle potential temperature ($T_P$) of 1500 °C and 1700 °C, along which melt was extracted to simulate adiabatic decompression melting. Modebox plots of melt fraction for $T_P$ = 1500 °C (B) and 1700 °C (C). Area of red rectangles represent the contribution of each 10 °C melt generation and extraction step to the crust, with the area sum of all rectangles proportional to the crustal thickness.